\documentclass[preprint, prd, floatfix, showpacs, superscriptaddress]{revtex4}
\usepackage{graphicx, epsfig, amsmath,amsthm,amssymb}

\begin{document}

\title{Gamma-ray Astronomy with Muons: Sensitivity of IceCube to PeVatrons in the Southern Sky}

\author{Francis Halzen}
\affiliation{Department of Physics, University of Wisconsin,
Madison, WI 53706, USA}

\author{Alexander Kappes}
\affiliation{Erlangen Centre for Astroparticle Physics, Universit\"at Erlangen-N\"urnberg, D-91058 Erlangen, Germany}

\author{ Aongus \'O\,Murchadha}
\affiliation{Department of Physics, University of Wisconsin,
Madison, WI 53706, USA}

\begin{abstract}
Northern hemisphere TeV gamma-ray observatories such as Milagro and Tibet AS$\gamma$ have demonstrated the importance of all-sky instruments by discovering previously unidentified sources that may be the PeVatrons producing cosmic rays up to the ``knee" in the cosmic ray spectrum. We evaluate the potential of IceCube to identify similar sources in the southern sky by detailing an analytic approach to determine fluxes of muons from TeV gamma-ray showers. We apply this approach to known gamma-ray sources such as supernova remnants. We find that, similar to Milagro, detection is possible in 10 years for point-like PeVatrons with fluxes stronger than several $10^{-11}$ particles $\rm{TeV^{-1}\, cm^{-2} \,s^{-1}}$. 
\end{abstract}

\maketitle

\section{Introduction and Summary of Results}

The IceCube neutrino detector is projected to observe 220 atmospheric muon neutrinos per day in a background of cosmic ray induced atmospheric muons traversing the detector at a rate of 1650\,Hz. Upgoing muon tracks initiated in or near the detector by neutrinos that traverse the Earth must be separated from a background of down-going cosmic ray muons, which dominates the neutrino signal by a factor $10^{6}$ at TeV energy. IceCube has demonstrated this capability with a partially deployed detector and has thus become the largest neutrino detector observing the northern sky as well as the largest instrument detecting muons from the southern sky\,\cite{Ahrens:2003ix}. In light of the recent discovery of sources of TeV gamma-rays that may be the PeVatrons powering the cosmic rays up to the ``knee" in the cosmic ray spectrum\,\cite{HalzKappOm_PeV}, it is important to revisit the potential of IceCube as a gamma-ray detector similar to the Milagro and Tibet AS$\gamma$ arrays but located in the Southern Hemisphere. This paper will focus on the possibility of identifying PeVatrons in the southern sky, as well as known bright gamma-ray sources such as the supernova remnant Vela Jr. 

Recently, the Milagro and Tibet AS$\gamma$ TeV gamma ray experiments have demonstrated the importance of constructing all-sky telescopes complementary to the more sensitive pointing air \c{C}erenkov gamma-ray telescopes, discovering a class of TeV sources that had not been previously distinguished by the pointing telescopes\,\cite{Abdo:2007ad, HESS1908}. The properties of these non-thermal sources are nevertheless striking, with spectra consistent with an $E^{-2}$ energy dependence possibly up to at least 100\,TeV. They are the first candidate sources to be associated with the PeVatrons that accelerate cosmic rays to the ``knee" at 3\,PeV. The sources cluster in the direction of the nearest spectral arms, consistent with the idea that cosmic rays are accelerated by the remnants of supernovae exploding in star-forming regions. If these sources are PeVatrons, the TeV gamma-rays are the decay products of neutral pions produced in the interaction of cosmic rays from the remnant with atoms in the interstellar medium near the source. Supernovae associated with molecular clouds are a common feature of associations of thousands of OB stars that exist throughout the Galactic plane and so the flux detected at Earth should be larger from the vicinity of known molecular clouds where the greater target density is the source of an enhanced TeV flux. Observations are consistent with this scenario: Milagro observes gamma-ray emission with an average energy of 10\,TeV over a broad patch of the sky in the direction of the star forming region in Cygnus as a background to individual sources, which could be associated either with molecular clouds or with the supernova remnants themselves.

We discuss here how it may be possible to identify similar sources in the Southern Hemisphere by operating IceCube as a gamma-ray telescope. The possibility of using muon detectors in general, and IceCube in particular, as a gamma-ray telescope has been extensively studied\,\cite{ SGH85, HHS86, Berez88, DHH89, bhatt2, HSY97, HH2003} and has been exploited to put limits on the TeV emission of a soft gamma-ray-repeater\,\cite{SGR, Achterberg:2006az}. Showers initiated in the atmosphere by TeV gamma rays from sources in the Southern Hemisphere produce muons that penetrate to the depth of the IceCube detector. Although few muons are produced in a gamma-ray shower relative to a hadronic shower, a directional source of TeV photons can nevertheless produce a statistically significant excess over the large background of muons produced by cosmic ray primaries. To operate as a gamma ray observatory, IceCube records the directions and energies of all ``background" muons to create a muon sky map of the southern sky, something that no other experiment is capable of at this time.   

Clearly IceCube cannot match the instantaneous sensitivity of the new generation of atmospheric \c{C}erenkov telescopes. The latter, however, are only capable of observing a several-degree patch of the sky on clear moonless nights. With a South Pole location, IceCube is unique in that it observes the same sky without interruption. It is also sensitive to sources overhead at the South Pole, a poorly studied portion of the southern sky. IceCube is the first large-scale detector in the Southern Hemisphere with the potential to detect TeV gamma-ray photons. The scientific potential of such all-sky TeV gamma-ray detectors has been clearly demonstrated by Milagro and Tibet AS$\gamma$.

Evaluating the capability of IceCube to observe Southern PeVatrons requires revisiting the estimates of the muon flux generated by photon showers. The sources of muons in a photon shower are threefold: the decay of mesons (pions and kaons) produced in photoproduction events by shower photons on nuclei in the atmosphere, muon pair production in the electric field of a nucleus,  and the production and decay of charm particles. At the energies considered in this paper muons of charm origin are few and their contribution is within the ``errors" associated with the first two mechanisms\,\cite{Berez88}. Section~\ref{sec:fluxes} discusses our analytical derivation of the muon flux created by a gamma-ray beam. Progress has been made possible by the appearance of simulation programs such as GEANT and CORSIKA. The EGS4 option in CORSIKA enables a full Monte Carlo simulation of the electromagnetic component of showers and provides detailed information for all electromagnetic particles. Apart from standard photon interactions like Compton scattering and $e^{+}e^{-}$ pair production CORSIKA also includes direct $\mu^{+} \mu^{-}$-pair production and photonuclear reactions with nuclei in the atmosphere\,\cite{CorsikaPhys}. We can compare this with the earlier analytic shower calculations to converge on more reliable estimates. While the linear shower calculations match the shower calculations for the rate of muons of pionic origin, we found that this is not the case for muon pair production. The commonly followed procedure of substituting the electron by the muon mass in the expression for electron pair production results in underestimating the high-energy cross-section by a factor 2 and so our reevaluation of the muon pair cross section following the formula used by GEANT results in an enhanced production of high energy muons in photon showers compared to CORSIKA (version 6.900). 

Our model does suffer from certain systematic defects intrinsic to an analytical approach. For example, the flux we derive is necessarily time-averaged, resulting in our inability to predict the rate of muon bundles due to the occurrence of several muon production events within a single gamma shower. However, examination of CORSIKA event rates shows that the rate of multiple production events is small and does not affect the final significance of the signal. The strength of our approach lies in its transparency: all physical parameters and cross-sections can be controlled and easily changed if necessary. As a practical matter the code written for this work runs in $<10\,\rm{s}$ on a personal computer, much faster than a full Monte Carlo, making this approach the correct one for a comprehensive initial examination of the possibility of detecting gamma-induced showers and a good first approximation to the full simulation that will be required to perform an analysis of real experimental data. 

\begin{figure}[htb]
\begin{center}
\epsfig{file=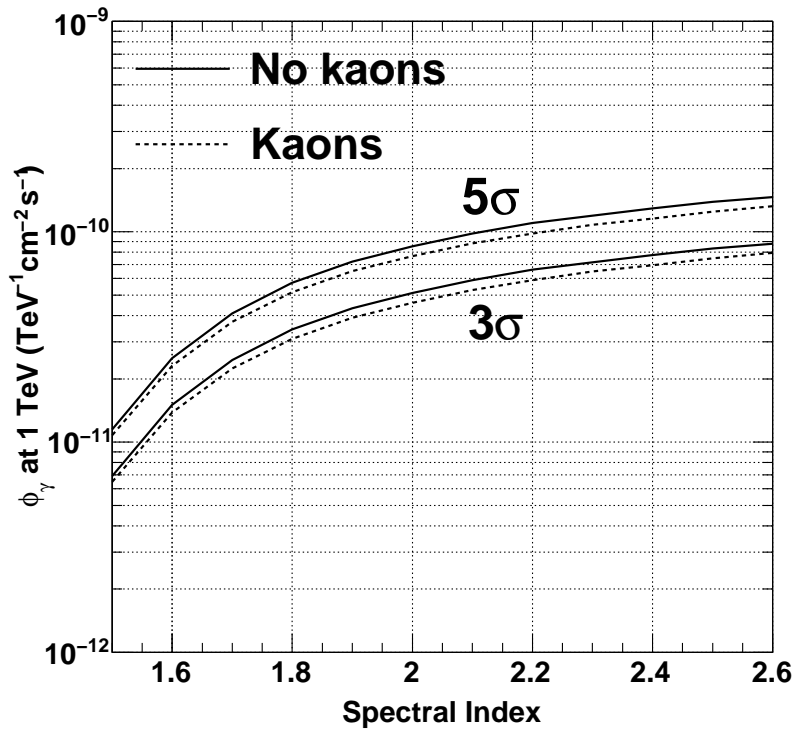, scale=0.55 }
\end{center}
\caption{Differential flux as a function of spectral index from a point-like PeVatron required to create an excess over background with a statistical significance of 3 and $5\,\sigma$ (probabilities of $1.3\times 10^{-3}$ and $2.9\times 10^{-7}$, respectively) after 10 years. The dashed lines include an estimated contribution from kaon production and decay (Section IIC). }
\label{fig:pevnorm}
\end{figure}

In Section~\ref{sec:sources} we discuss the primary sources of cosmic gamma-rays, in particular the PeVatrons producing pionic gamma rays, whose spectrum extends to several hundred TeV without cut-off, in interactions with the interstellar medium. By straightforward energetics arguments\,\cite{Halzen:2007ah} one anticipates the TeV flux from a source at a nominal distance of 1\,kpc to be in the range
\begin{equation}
{E {dN_{\rm events}\over dE}}\,({ >}1\,\rm{TeV}) = {L_\gamma \over 4\pi d^2} \simeq 10^{-12}-10^{-11}  \left({\rm photons\over \rm cm^2\,s}\right) \left({W\over \rm 10^{50}\,erg}\right) \left({n\over \rm 1\,cm^{-3}}\right) \left({d\over \rm 1\,kpc}\right)^{-2}.
\end{equation}
Such sources must emerge in an all-sky TeV gamma ray survey performed with an instrument with the sensitivity of the Milagro experiment. Based on observations by Milagro and H.E.S.S., it has been argued that one Pevatron, MGRO\,J1908+06, has likely been identified among six candidates in the current sky map\,\cite{HalzKappOm_PeV}. Our main conclusion (Figure~\ref{fig:pevnorm}) is that a generic point-like PeVatron with a normalization at 1 TeV of $3-8\times10^{-11}\,\rm{TeV^{-1}\,cm^{-2}\, s^{-1}}$ extending to 300\,TeV may be observed with a significance of $3-5\, \sigma$ in 10 years by counting the total detected number of muons over the full energy range. We also emphasize the possibility of exploiting the complementarity between all-sky and pointed telescopes that led to the final detailed observation and flux measurements of J1908+06: once a significant excess has been observed in IceCube, a more sensitive pointed instrument can devote observation time to that specific region of the sky, with luck confirming the presence of a source and measuring the spectrum of gamma-rays. 

\section{Muon Fluxes From Gamma Showers}
\label{sec:fluxes}

\subsection{Gamma-ray cascades in the atmosphere }

The $\gamma$-rays that reach the Earth from the source will interact with an atom of the atmosphere producing an electron-positron pair, which will then via bremsstrahlung emit photons and create an electromagnetic cascade. Since the photons of the cascade can create muons that are detectable in IceCube, we must first find the spectrum of $\gamma$-rays at some slant depth $t$ in the atmosphere $\gamma\left(E, t\right)$ created by the initial source flux.  If our initial (differential in energy) spectrum can be represented by a power-law $\gamma\left(E, t=0\right)= K_{\gamma}E^{-(s+1)}$, cascade theory under Approximation A\,\cite{GreisenRossi} tells us that the spectrum after cascading is a power-law of the same spectral index: 

\begin{equation}
\label{eq:gammagen}
\gamma\left(E, t\right) = K_{\gamma}E^{-(s+1)}\frac{(\sigma_{0} + \lambda_{1})(\sigma_{0} + \lambda_{2})}{\lambda_{2}-\lambda_{1}}  \left[ \frac{e^{\lambda_{1}t}}{\sigma_{0} + \lambda_{1}}  - \frac{e^{\lambda_{2}t}}{\sigma_{0} + \lambda_{2}}  \right] \equiv \gamma\left(E, t=0\right)\gamma_{2}\left( t \right),
\end{equation}
$\lambda_{1,2}$ are the scale lengths of the shower growth and attenuation in the atmosphere and $\sigma_{0}\sim 7/9$ is the probability of $e^{+}e^{-}$ pair production per radiation length. The $\lambda$'s are dependent on the integral spectral index $s$ and are given in tabulated form in Ref.~\cite{GreisenRossi}. Conventionally, $t$ is in units of radiation (bremsstrahlung) length and so the parameters $\lambda_{i}$ and $\sigma_{0}$ have units of (radiation length)$^{-1}$. The most important special case is $s=1$, because $\lambda_{1}(s=1)=0$ and $\lambda_{2}(s=1)<0$. As a result, an incident differential flux that goes as $E^{-2}$ will create a cascade that maximizes in the atmosphere after several radiation lengths, without any subsequent attenuation:

\begin{equation}
\label{eq:gammae-2}
\gamma\left(E, t\right) \rightarrow 0.567\, \gamma\left(E, t=0\right). 
\end{equation}

\subsection{Pion production}
The principal channel for gamma-ray induced muons is by the photoproduction of pions on the nuclei of the atmosphere. Drees et al. derive a closed-form analytical solution for this flux using the linear cascade equations, assuming $E^{-2}$\,\cite{DHH89}. In the following section, we generalize their method to other spectral indices, essentially starting with Eq.~\ref{eq:gammagen} rather than with Eq.~\ref{eq:gammae-2} for the gamma-ray flux. Due to the explicit dependence on $t$ in Eq.~\ref{eq:gammagen}, the solution for a general spectral index cannot be expressed in a closed form and must be represented by an integral. However, this integral is not difficult to compute numerically and allows us to avoid the assumptions regarding depth that are required to have no final dependence on $t$, even for $E^{-2}$. For a general derivation of the linear cascade equation and its simplification under Approximation A we recommend the discussion in Ref.~\cite{gaisser90}. 

We begin with the ansatz that the pion spectrum (differential in energy) can be factorized as 
\begin{equation}
\pi(E, t) = \gamma\left(E, t=0\right) \pi_{2}(E,t).
\end{equation}

Where $\lambda_{\gamma(\pi) A}$ and $\sigma_{\gamma(\pi) A}$ are respectively the interaction length and cross-section of photons (pions) on nuclei of mass number $A$ (in this case air nuclei with $\langle A \rangle = 14.5$) and using the standard definitions of spectrum-weighted moment $z_{ij}$, effective interaction length $\Lambda_{i}$, and pion decay length $d_{\pi}(t)$ \,\cite{gaisser90}, we can reduce the linear cascade equation

\begin{equation}
\label{eq:pioncascade}
\frac{d\pi}{dt}\left(E,t \right) = -\left[ \frac{1}{\lambda_{\pi A}} + \frac{1}{d_{\pi}(t)} \right] \pi\left(E,t \right)\, +\int_{0}^{1}\frac{dx}{x}\frac{\gamma(E/x, t)}{\lambda_{\gamma A}}\frac{1}{\sigma_{\gamma A}}\frac{d\sigma_{\gamma\rightarrow\pi}(x)}{dx}\,+ \int_{0}^{1}\frac{dx}{x}\frac{\pi(E/x, t)}{\lambda_{\pi A}}\frac{1}{\sigma_{\pi A}}\frac{d\sigma_{\pi\rightarrow\pi}(x)}{dx}
\end{equation}
to 

\begin{equation}
\label{eq:pionhediffeq}
\frac{d\pi_{2}}{dt}(E,t)  =  -\left[ \frac{1}{\Lambda_{\pi}}+ \frac{1}{d_{\pi}(t)} \right] \pi_{2}(E,t)\, + \frac{z_{\gamma \pi}}{\lambda_{\gamma A}}\gamma_{2}(t).
\end{equation}
The first term in Eq.~\ref{eq:pioncascade} represents the loss of pions due to interactions and decay, and the second term represents the production of pions in photonuclear interactions. The production of lower-energy pions in interactions of pions with air, the third term in Eq.~\ref{eq:pioncascade}, has the effect of reducing the loss of pions due to interactions and is taken into account by using the effective interaction length $\Lambda_{\pi}$ in Eq.~\ref{eq:pionhediffeq} which is larger than the normal interaction length $\lambda_{\pi}$.  We assume that the inclusive differential cross-sections depend only on the fraction $x$ of energy transferred from the parent particle (Feynman scaling), and that the interaction lengths $\lambda_{\gamma(\pi) A}$ are constant over the energies considered here.

We solve this differential equation in two regimes: high energy where pion interactions dominate over decay ($\Lambda_{\pi}\ll d_{\pi}$), and low energy where we can neglect pion interactions altogether ($d_{\pi}\ll \Lambda_{\pi}$). The energy scale is set by the pion decay energy constant $\epsilon_{\pi}=115$ GeV in the decay length $d_{\pi} = E\,t \cos\theta/\epsilon_{\pi}$, where $\theta$ is the zenith angle of the incident gamma-ray flux. At high energy, where we can solve Eq.~\ref{eq:pionhediffeq} by hand, 

\begin{equation}
\pi_{2}^{HE}(t) =\frac{z_{\gamma \pi}}{\lambda_{\gamma A}}\frac{(\sigma_{0} + \lambda_{1})(\sigma_{0} + \lambda_{2})}{\lambda_{2}-\lambda_{1}}\left[    \frac{e^{\lambda_{1}t}-e^{-t/\Lambda_{\pi}}}{(\sigma_{0} + \lambda_{1})(\lambda_{1} + \frac{1}{\Lambda_{\pi}})}             -\frac{e^{\lambda_{2}t}-e^{-t/\Lambda_{\pi}}}{(\sigma_{0} + \lambda_{2})(\lambda_{2} + \frac{1}{\Lambda_{\pi}})}                           \right]. 
\end{equation}
At low energy,

\begin{equation}
\label{eq:le_pi}
\pi_{2}^{LE}(E, t) = \frac{z_{\gamma \pi}}{\lambda_{\gamma A}} \frac{(\sigma_{0} + \lambda_{1})(\sigma_{0} + \lambda_{2})}{\lambda_{2}-\lambda_{1}} \int_{0}^{t}dt' \left(  \frac{t'}{t}  \right)^{\delta} \left[  \frac{e^{\lambda_{1}t'}}{\sigma_{0} + \lambda_{1}} -   \frac{e^{\lambda_{2}t'}}{\sigma_{0} + \lambda_{2}}\right],
\end{equation}
where $\delta = t/d_{\pi}$.

For spectral indices such that $\lambda_{1},\,\lambda_{2} < 0$ the integral of Eq.~\ref{eq:le_pi} can be expressed as the lower incomplete gamma function. However, $\lambda_{1}>0$ for spectra harder than $E^{-2}$, and moreover at low energies ($\delta\gg 1$) and small depths the integrand will be so close to zero that the standard numerical implementations of the gamma function will fail.  For these reasons we instead expand the exponential factor in the integrand as a series and perform the integration term-by-term:

\begin{equation}
\int_{0}^{t}dt' \left(  \frac{t'}{t}  \right)^{\delta} \frac{ e^{\lambda_{i}t'}}{\sigma_{0} + \lambda_{i}} \approx \frac{1}{\sigma_{0} + \lambda_{i}}\sum_{j=1}^{100} \frac{\lambda_{i}^{j-1} \, t^{j}}{(j-1)!(\delta + j)}.
 \end{equation}
Due to the factorial in the denominator the series converges quickly and so we terminate the summation after 100 terms. The final pion flux is then an interpolation of the high- and low-energy limiting forms. However, since we do not have analytical expressions for both, smoothly transitioning from one limit to another is difficult and we take simply

\begin{equation}
\pi(E, t) = \gamma\left(E, 0\right) \min \left[  \pi_{2}^{HE}(t), \pi_{2}^{LE}(E, t) \right]. 
\end{equation}
Comparing this to the smooth interpolation in Ref.~\cite{DHH89}, we see that any discrepancy is at the transition from low- to high-energy behavior. This occurs at energies around $\epsilon_{\pi}/\cos\theta$ which is too low to produce detectable muons. Assuming no muon decay or energy loss in the atmosphere, standard 2-body decay kinematics\,\cite{gaisser90} give the muon flux at the surface

\begin{equation}
\frac{dN_{\mu}}{dE} = \int_{0}^{t_{\rm{max}}} \!dt \,B_{\mu \pi} \!\int_{E}^{E/r} \frac{dE'}{(1-r)E'} \,\frac{\pi(E', t)}{d_{\pi}(t)},
\end{equation}
where $r = (m_{\mu}/m_{\pi})^{2}$ and $B_{\mu \pi} = 1$ is the number of muons per decaying pion. The upper cutoff on the depth integral $t_{\rm{max}}$ arises from the fact that the gamma-ray flux does not extend to arbitrarily high energies, so that at some depth in the cascade there should remain no photons with sufficient energy to produce a muon of energy $E$. Following the `Heitler model' argument of Ref.~\cite{DHH89} we take 

\begin{equation}
t_{\rm{max}} = \lambda_{H}\ln \left[ \frac{E_{\rm{max}}\,\langle x \rangle_{\gamma\rightarrow\mu}}{E}    \right],
\end{equation}
where we define the effective Heitler cascade length $\lambda_{H} = 8/7$, the average of the bremsstrahlung and pair production lengths. For pionic decays the fraction of gamma-ray energy that goes into the final muon is $\langle x \rangle_{\gamma\rightarrow\mu} = 0.25$. This form for $t_{\rm{max}}$ causes the final muon flux to depend approximately logarithmically on $E_{\rm{max}}$. For very low-energy muons, if we neglect muon decay the limiting factor is the physical extent of the atmosphere rather than the depletion of high-energy photons in the cascade, so $t_{\rm{max}}$ cannot be greater than (680/$\cos\theta)~\rm{g/cm}^{2}$ (18.3/$\cos\theta$ radiation lengths), the atmospheric depth of the South Pole ice surface. However, this depth is so large that it does not affect the muon flux that reaches the detector for any reasonable $E_{\rm{max}}$: the muon energy that corresponds to $t_{\rm{max}}=18.3$ is given by $E = E_{\rm{max}}\times(2.8\times10^{-8})$ and only the production of muons with less energy is suppressed due to the ice surface.  Since the initial particle is a gamma-ray, taking $\lambda_{H} = 8/7$ underestimates the depth integral since the first electromagnetic cascade length dominates with $\lambda_{e^{+}e^{-}} = 9/7$. Comparison with the Monte Carlo results of Ref.~\cite{HHS86} and CORSIKA reveals that the analytic approach suffers from a systematic underestimation on the order of 20\%, so in this work we multiply the muon fluxes from mesons by a factor 1.2. 

We take the particle physics parameters used here from Ref.~\cite{DHH89}. We convert all lengths to radiation lengths, assuming the radiation length in air to be $37.1\,\rm{g/cm^{2}}$. We assume the photoproduction cross-section $\sigma_{\gamma A} = A\,\sigma_{\gamma N} = 14.5\times 100\,\rm{\mu b}$\,\cite{SGH85}. We obtain interaction lengths in air via
\begin{equation}
\lambda= 2.4\times10^{4}~\left( \frac{\rm{1\,mb}}{\sigma} \right) \left[\rm{g/cm^{2}} \right] .
\end{equation} 
 For the pion interactions $\Lambda_{\pi} = \lambda_{\pi}/(1-z_{\pi\pi}) = 173\,\rm{g/cm^{2}}$. Finally we assume that far from the resonance region equal numbers of positive, negative and neutral pions are produced which gives $z_{\gamma \pi} = 2/3$ for an incident $E^{-2}$ (s=1) spectrum, and that $z_{\gamma \pi}$ and $z_{\pi \pi}$ vary sufficiently little that we can take the $s=1$ value for all spectral indices of interest.

\subsection{Kaon production}

At high energies, the production of muons from kaon decay becomes significant. A precise treatment of kaons would complicate the determination of muon fluxes greatly, however. Kaons decay to pions, and pions can create kaons in interactions with atoms in the air, requiring not only an extra cascade equation for kaons but also additional source terms in the cascade equations that couple the meson fluxes. Fortunately, the coupling is weak: the z-factor for kaon production in pion interactions is of the order of 0.01 \,\cite{lipari93}, and while the branching ratio for hadronic kaon decays is significant \,\cite{PDG}, as is pion production in kaon interactions, the kaon flux is sufficiently small that its contribution to the pion flux can be neglected. As a result, we consider the fluxes of pions and kaons to be independent, neglecting the production both of pions by kaons and of kaons by pions. This allows us to re-use the pion formalism detailed above to estimate the kaon flux, substituting parameters where appropriate. We estimate the kaon production cross-section for energies far above threshold as $\sigma_{\gamma N \rightarrow K} = \sigma_{\gamma N \rightarrow \pi}\, (m_{\pi}/m_{K})^{2}$. This gives $\sigma_{\gamma N \rightarrow K} \sim 8\, \mu b$, a reasonable estimate since that is the order of the total cross-section near threshold \,\cite{kaons1, kaons2} and we expect a logarithmic increase in cross-section with energy to approximately counteract the fall in cross-section above the resonant threshold region. As with pions, we assume that 2/3 of the produced kaons are charged ($z_{\gamma N \rightarrow K} = 2/3$) and we neglect the neutral kaons. This results in underestimating the final muon flux since the neutral kaons typically decay either hadronically into pions or semileptonically into leptons and charged pions, but even were these to be treated here the uncertainties in the cross-section preclude estimating the kaon-induced muon flux with any great precision. We also neglect the hadronic decays of the charged kaon, considering only the 63.5\% that decay to a muon and a muon neutrino: $K^{\pm}\rightarrow \mu^{\pm}+\nu_{\mu}$. Therefore $B_{\mu K} = 0.635$. The scale energy of kaon decay in air is $\epsilon_{K} = 850$ GeV and we take the effective interaction length $\Lambda_{K} = 200\,\rm{g/cm^{2}} $. For kaon decays $\langle x \rangle_{\gamma\rightarrow\mu} = 0.17$ assuming an equal average fraction of gamma-ray energy into kaons as into pions. Due to the large uncertainties in the kaonic muon flux all event rates and statistical significances in this work are given both with and without the kaon contribution.

\subsection{Muon pair production}

\subsubsection{Cross-section}
One of the primary mechanisms by which an electromagnetic cascade proceeds is by the pair production of electrons by the shower photons. As a result, electron pair production has been understood since the earliest days of cosmic-ray physics\,\cite{BetheHeitler} but its generalization to more massive leptons has in many cases been flawed. In this section we describe the physics of pair production of massive leptons by an energetic photon, and discuss and correct the standard errors that have found their way into the literature and hence into standard software packages such as the Monte Carlo event generator CORSIKA (version 6.900). 

\begin{figure}[hbt]
\begin{center}
\epsfig{file=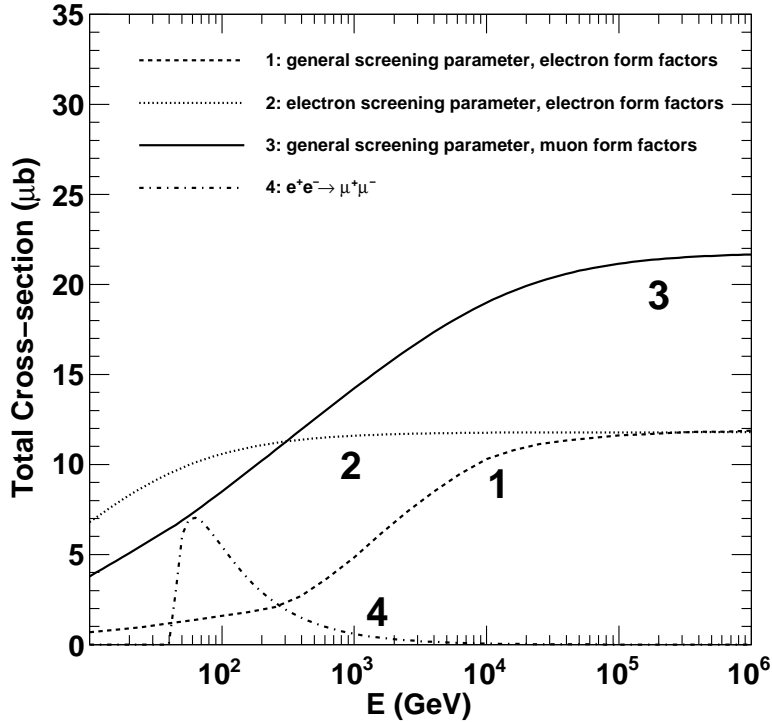, scale=0.55 }
\end{center}
\caption{Muon pair production cross-section on air (Z=7.2): 1 has the correct screening but uses electron pair production atomic form factors\,\cite{SV85}; 2 is the Bethe-Heitler electron pair cross-section with $m=m_{\mu}$\,\cite{MotzOlsenKoch}; 3 is the correct cross-section~\cite{GEANT} with inelastic corrections~\cite{KelKokPet};  finally muon pair production via positron annihilation with atomic electrons (see Section~\ref{sec:elecprocs} below). }
\label{fig:mupr}
\end{figure}

Pair production occurs when an incident photon interacts with a photon from the electric field of a nucleus, producing a pair of leptons. Since a single photon cannot produce a pair of massive particles without violating the conservation of 4-momentum, a second photon is necessary, transferring the required momentum from the nucleus. As the energy of the primary photon increases, the minimum momentum that the second photon must have for the pair production to occur decreases. As a result, the maximum distance from the nucleus at which the primary photon can still initiate pair production increases with energy and therefore we expect the cross-section to similarly increase. If the nucleus were a bare source of electric field we would expect the cross-section to increase without bound, as at extremely high energy the minimum momentum transfer necessary would be so small as to be possible at very large distances from the nucleus. However, nuclei are not electric monopoles out to arbitrarily large impact parameters. At a certain radius the nuclear electric field becomes `screened' by the opposite electric field of the atomic electrons and at very large distances the atom is essentially electrically neutral. This decrease in effective electric charge beyond a certain distance from the nucleus cuts off the logarithmic increase of the cross-section and leads to the cross-section reaching an asymptotic maximum at high energy. Therefore, accurately predicting the cross-section requires estimating the maximum distance from the nucleus that pair production of a gamma-ray of energy $E_{\gamma}$ into two leptons of energy $E_{+}$ and $E_{-}$ can occur, and then determining the effective nuclear charge visible at that distance and hence the probability of the emission of a virtual photon of the required momentum. The distance estimation is generally expressed by the screening parameter $\delta\sim r_{Z}/R_{\rm{max}}$, essentially the ratio of the radius of the atomic cloud $r_{Z}$ and the maximum impact parameter, $R_{\rm{max}}$, to obtain the minimum necessary momentum transfer. From the conservation of energy and momentum and assuming energies large compared to the mass of the pair particles $m_{l}$, we find the minimum necessary momentum transfer from the nucleus to be
\begin{equation}
q_{\rm{min}} = \frac{m_{l}^{2}\,E_{\gamma}}{2E_{+}E_{-}}.
\end{equation} 
Therefore, the maximum distance from the nucleus at which this momentum transfer can occur is
\begin{equation}
R_{\rm{max}} \sim \frac{\hbar}{q_{\rm{min}}}.
\end{equation} 
Following the Thomas-Fermi model, the effective radius of an atom of atomic number $Z$ is
\begin{equation}
r_{\rm{Z}} = \frac{\hbar^{2}}{m_{e}\,e^{2}}\,Z^{-1/3}.
\end{equation} 
Therefore, the screening parameter is of the form
\begin{equation}  
  \delta \sim \frac{r_{Z}}{R_{\rm{max}}}=\frac{m_{l}^{2}\,E_{\gamma}}{2E_{+} E_{-}(\alpha\, m_{e}\, Z^{1/3} ) }.
\end{equation}
Note that if in describing electron pair production we take $m_{l} = m_{e}$ and cancel the mass factor in the denominator, the final expression will no longer be generalizable to other pair particle masses. $\delta$ will be too small by a factor $m_{e}/m_{l}$, resulting in the interaction reaching the  full screening regime at overly low energies. For muon pairs on nitrogen, $\delta = 1$ corresponds to a gamma-ray with energy $\sim100$\,TeV, which is therefore the approximate energy at which we expect screening to begin to saturate the cross-section. We see this by comparing in Fig.~\ref{fig:mupr} curve 1, which uses the correct general expression for $\delta$, to curve 2, which uses the reduced form of $\delta$ correct for electron screening with the substitution $m=m_{\mu}$.

The second common error comes from the fact that the form factor of the atom must be integrated over the transferred momentum, where the upper limit is $\sim m_{l}$.  Therefore, the correct result for electron pair production is not generalizable to heavier particles and the limit at high energy will be too small by a factor of order $\ln(m_{l}/m_{e})$ (curves 1 and 2 in Fig.~\ref{fig:mupr})\,\cite{Berez88}. In this work, we take the muon pair production formula of Ref.~\cite{GEANT}, adding also the inelastic contribution using the form factors from Ref.~\cite{KelKokPet} (curve 3 in Fig.~\ref{fig:mupr}). We note that the current version of the standard cosmic-ray Monte Carlo event generator CORSIKA (version 6.900) has been written to reproduce the cross-section of Ref.~\cite{SV85} (curve 1), whose asymptotic value is a factor $\sim2$ smaller than the correct value, and that this will be corrected in future versions\,\cite{Heckprivate}.  

\subsubsection{Muon flux from pair production}
The spectrum of pair muons is derived from the cascade equation

\begin{equation}
\frac{d\mu}{dt} = 2\int_{0}^{1}\frac{dx}{x}\frac{\gamma(E_{\mu}/x, t)}{\lambda_{\mu \mu}}\frac{1}{\sigma_{\mu \mu}}\frac{d\sigma_{\gamma\rightarrow\mu\mu}}{dx}\left(x, E_{\gamma} = \frac{E_{\mu}}{x}\right),
\end{equation}
giving us, including a factor 2 for the two muons produced, and taking the cross-section in millibarns (mb) and depth $t$ in radiation lengths, 

\begin{equation}
\label{eq:muprexact}
\frac{dN_{\mu}}{dE_{\mu}} = \frac{2\,\gamma_{0}(E_{\mu})\,\lambda_{\rm{rad}}(\rm{g/cm}^{2})}{2.4\times10^{4}}\int^{t_{\rm{max}}}_{0} dt\, \gamma_{2}(t, s)\int^{1}_{0}dx \, x^{s}\,\frac{d\sigma_{\mu\mu}}{dx}\left(x, \frac{E_{\mu}}{x}\right). 
\end{equation}
Equation~\ref{eq:muprexact} assumes that both muons in the pair are separately measureable. In practice, however, detectors such as IceCube do not have the spatial resolution to distinguish between the two muons of the pair and will instead reconstruct a single muon with the total energy of the pair, the energy of the parent gamma-ray. The pair spectrum will therefore be

\begin{equation}
\frac{dN_{\rm{pair}}}{dE} = \frac{\gamma_{0}(E) }{2.4\times10^{4}} \left(\frac{\lambda_{\rm{rad}}}{\rm{g/cm}^{2}}\right)\left(\frac{\sigma_{\mu\mu} (E)}{\rm{mb}}\right)\int^{t_{\rm{max}}}_{0} dt\, \gamma_{2}(t, s)
\end{equation}
where

\begin{equation}
\sigma_{\mu\mu} (E) = \int^{1}_{0} dx \, \frac{d\sigma}{dx}(x, E)
\end{equation}
is the total cross-section for a gamma-ray of energy $E$ to make a muon pair.

\subsection{Muon production via electrons}
\label{sec:elecprocs}

Both processes outlined above, muon pair production and meson photoproduction, have analogues with an incident electron rather than an incident photon. It is possible for a cascade positron to annihilate an atomic electron, producing a muon pair $e^{+} + e^{-} \rightarrow \mu^{+} + \mu^{-}$ and it is also possible for a cascade electron or positron to undergo an electronuclear interaction by producing a virtual photon, possibly resulting in the creation of a meson $e^{\pm} + A \rightarrow e^{\pm} + \gamma^{*}+ A \rightarrow e^{\pm} + A + \pi/K$.

 In practice, however, both these processes are negligible. The peak cross-section for muon pair production from electron pair annihilation is $\sim1 \,\mu$b per electron\,\cite{AkhBer}, giving $\sim Z \,\mu$b per atom (Fig.~\ref{fig:mupr}). Away from this peak, which occurs at a positron energy of 61\,GeV, the cross-section falls off rapidly and is $0.2\,Z \,\mu$b at 500\,GeV positron energy and is therefore essentially zero at energies that are high enough to produce muons with the $>$\,600\,GeV of energy that is needed to reach the detector through the ice overburden. By contrast, the cross-section for the electronuclear interaction is approximately 80\% of the photonuclear interaction cross-section but the average fraction of energy that the virtual photon carries is only about 5\% of the incident electron energy\,\cite{MuonEas}. As a result, nuclear interactions of charged particles are an important energy loss mechanism without being a significant source of secondary particles. A necessary caveat is that photonuclear interactions are not theoretically well-understood and the existing formulas for lepton-nucleus interactions are largely phenomenological descriptions designed to reproduce the data for muon energy loss in matter. The above numbers for electronuclear energy loss are derived from a muon energy loss formula using the substitution $m_{\mu}\rightarrow m_{e}$, which is clearly a crude approximation. However, based on the example of muon pair production from a real photon, we do not expect our error due to this simple substitution to be greater than a factor $\sim2$ and we can reasonably conclude that electronuclear interactions can be neglected in this analysis. 

\section{Muon Propagation and Energy Loss}

Neglecting muon energy loss and decay in the atmosphere, we treat muon energy loss in the ice above the detector by taking the standard average energy loss formula:

\begin{equation}
\left< \frac{dE}{dx} \right> = -a - bE
\end{equation} 
with $a =  2.59 \times 10^{-6} \,\rm{TeV\,(g/cm^{2})^{-1}}$ and $b = 3.63 \times10^{-6} \,\rm{(g/cm^{2})^{-1}}$ in ice\,\cite{Chirkin04}. The general solution for final muon energy $E_{f}$ at depth $R$ given initial muon energy $E_{i}$ is 

\begin{equation}
\left<  E_{f}\left( R\right) \right> = \left(  E_{i} + a/b \right) e^{-bR} - a/b
\end{equation}
The flux at depth $R$ is then, for $E_{i} = (E_{f} + a/b)e^{bR} - a/b$, 

\begin{equation}
\frac{dN}{dE_{f}}(R, E_{f}) = \frac{dN}{dE_{i}}\left( E_{i} \right) e^{bR}. 
\end{equation}
Assuming IceCube to be at a vertical depth of 1450 m of ice with density $\rho = 0.92 \,\rm{g/cm^{3}}$, we find that for a vertical muon to arrive at the detector with energy 100\,GeV, the surface energy must be 607\,GeV. Of course, as the range goes as $d_{\rm{vert}}/\cos\theta$, the necessary surface energy will increase with increasing zenith angle.  

In the case of muon pairs, where due to the lack of spatial resolution we consider the flux of pairs rather than the flux of muons, the energy loss of the pair will be the sum of the energy losses of the individual muons. Although the pair muons each take on average half of the gamma-ray energy, the distribution is flat at low energies and has a minimum at $x=0.5$ at high energies. Therefore taking a simple average is a poor approximation of the energy splitting and will overestimate the energy losses of the pair. We consider the average energy fraction taken by the higher-energy muon: at 1 TeV $\langle x \rangle_{\rm{HE}}= 0.75$ and varies by only 4\% between 100 GeV and  1 PeV. The final energy of a muon pair of initial energy $E$ is therefore taken to be the sum of the final energies of two muons with initial energies $0.75\,E$ and $0.25\,E$. 

We assume that the angle between the directions of the muons and the direction of the parent gamma-ray is sufficiently small so that the distance between the muons on arrival at the detector is negligible compared to the spacing of the optical modules. While this is true at high energy, at low energies or large energy asymmetry between the muons one or both muons may miss the detector entirely. We neglect this possibility here as muons with a large angle to the gamma ray will be of low enough energy that they would not trigger the detector even if they were collinear to the gamma-ray.

\section{Detector Background and Effective Area}

Finding sources of cosmic gamma-rays will depend on whether or not the signal in muons is sufficiently large to be detected against the background of cosmic-ray induced muons. As a result, we have to determine the spectrum and event rates of atmospheric muons in IceCube. We must also determine IceCube's response to downgoing high-energy muons to determine the signal event rates. To do this, we have used the atmospheric muon event rates in IceCube\,\cite{ICprivate} and have extracted effective areas from them. Given these effective areas we can then find the event rates from gamma-ray sources. We do not consider extensions to IceCube-80 such as the six-string cluster ``DeepCore,'' as the additional optical modules are concentrated in the lower part of the detector and are therefore unlikely to significantly affect the effective area for low-energy downgoing muons. 

\begin{figure}[h]
\begin{center}
\epsfig{file=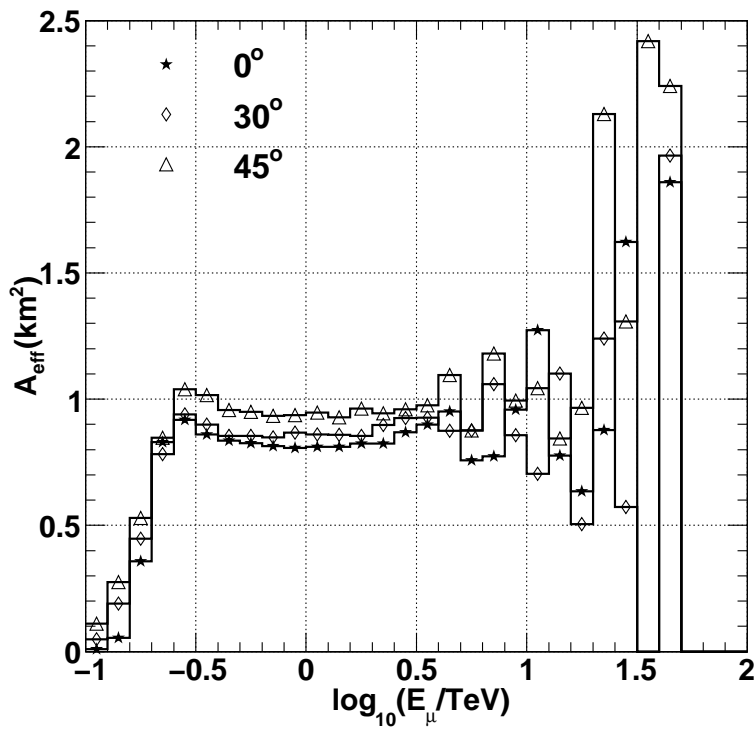, scale=0.55 }
\end{center}
\caption{Downgoing muon effective area for three zenith angles. The fluctuations at high energy are due to fluctuations in the generated background event rate and do not affect the significance of signal event rates at lower energies. }
\label{fig:aeff}
\end{figure}

The integral event rate $N(>\!\!\!E)$ is found by the convolution of the energy-dependent effective area with the muon flux at the detector, where T is the time interval:

\begin{equation}
N_{\mu}(>\!\!E) = T\!\int_{E}\, A_{\rm{eff}}(E_{\mu})\, \frac{dN_{\mu}}{dE_{\mu}}\left(E_{\mu}\right) \,dE_{\mu}
\label{eq:muareaconv}
\end{equation} 
Therefore, given an atmospheric muon event rate we have to make an assumption about the cosmic-ray-induced muon flux to find the effective area. We have used the analytical approximation given in Chapter 6 of Ref.~\cite{gaisser90} after propagation through the ice, and deconvolved the downgoing single muon effective area for three choices of zenith angle (Figure~\ref{fig:aeff}). Due to the event rate being the output of a Monte Carlo simulation, it is not a smooth function, particularly at high energies where it is susceptible to fluctuations due to poor statistics and as a result, the derived effective area has corresponding unphysical ``fluctuations.'' These do not affect our final conclusions as there are not sufficient events at those energies to affect the overall detectability of sources. At low energies the function matches our expectations for the downgoing muon effective area--an initial region of increasing area due to rising trigger efficiency followed by a region of approximately constant effective area, somewhat less than the standard $1\,\rm{km}^{2}$ due to the fact that IceCube's optical modules point down, reducing their sensitivity to light from downgoing muons. The `trough' at intermediate energies is most likely due to the muon multiplicity of proton showers increasing with energy, which the time-averaged analytical approximation must neglect. 


We assume that the muons contributing to the effective area have an angular resolution of $0.5^{\circ}$, and find the radius of our circular search bin to determine the background flux magnitude from 
\begin{equation}
(\rm{bin~radius}) = 1.6\times\sqrt{(\rm{angular~resolution})^{2} + (\rm{source~radius})^{2} }.
\end{equation}
This search bin will contain 72\% of the total signal events\,\cite{Kappes:2006fg}. The final significance of the excess above background in the search bin due to the signal is estimated using 

\begin{equation}
n_{\sigma}= \frac{0.72\,N_{\rm{sig}}}{\sqrt{N_{\rm{bkg}}}} ~(\rm{standard~deviations}).
\end{equation} 
We use the number of background events $N_{\rm{bkg}}$ in the denominator rather than the more conservative $N_{\rm{bkg}}+N_{\rm{sig}}$, due to the fact that IceCube can measure the mean background in the bin at any time by averaging over the total events from the declination band of the sky at the source zenith.   

\section{Sources of Cosmic Gamma-rays}
\label{sec:sources}

\subsection{PeVatrons}

\begin{figure}[p]
\begin{center}
\epsfig{file=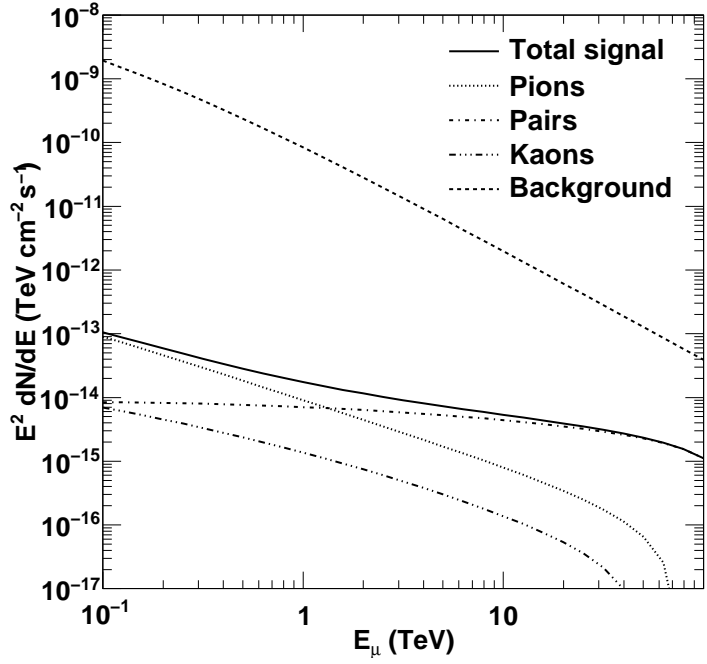, scale=0.5 }
\end{center}
\caption{Differential flux of muons at the surface from a point-like PeVatron with $E^{-1.6}$. }
\label{fig:pevflux16}
\end{figure}

\begin{figure}[hp]
\begin{center}
\epsfig{file=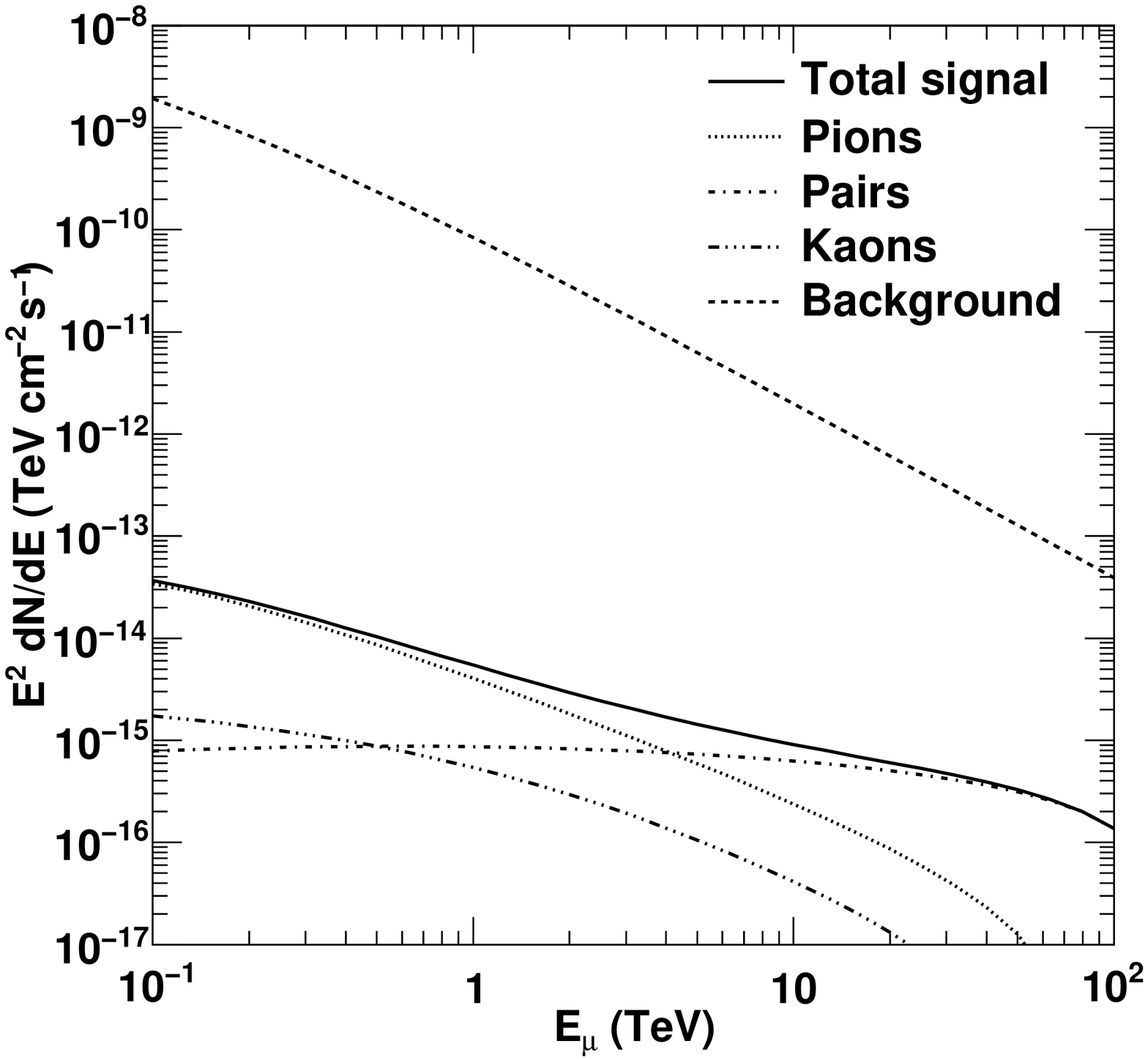, scale=0.5 }
\end{center}
\caption{Differential flux of muons at the surface from a point-like PeVatron with $E^{-2}$. }
\label{fig:pevflux2}
\end{figure}

\begin{figure}[hp]
\begin{center}
\epsfig{file=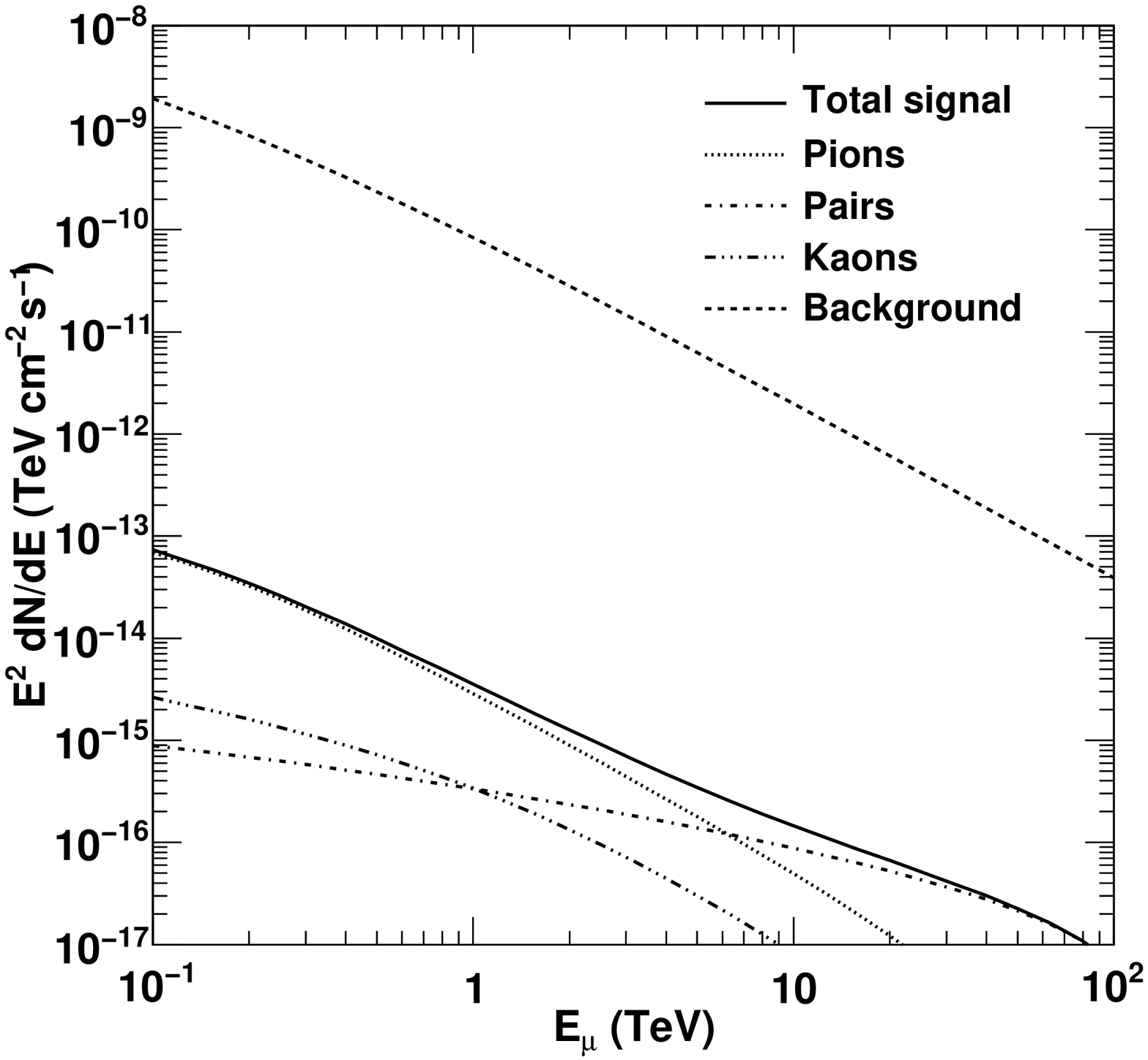, scale=0.5 }
\end{center}
\caption{Differential flux of muons at the surface from a point-like PeVatron with $E^{-2.6}$. }
\label{fig:pevflux26}
\end{figure}

\begin{figure}[hp]
\begin{center}
\epsfig{file=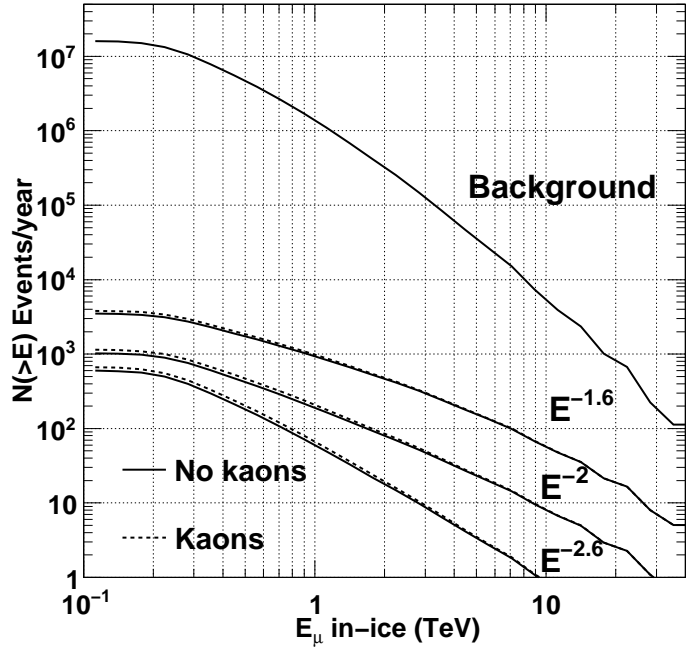, scale=0.5 }
\end{center}
\caption{Integral event rate per year from a point-like PeVatron for $E^{-1.6}$, $E^{-2}$ and $E^{-2.6}$. }
\label{fig:pevev}
\end{figure}

\begin{figure}[hp]
\begin{center}
\epsfig{file=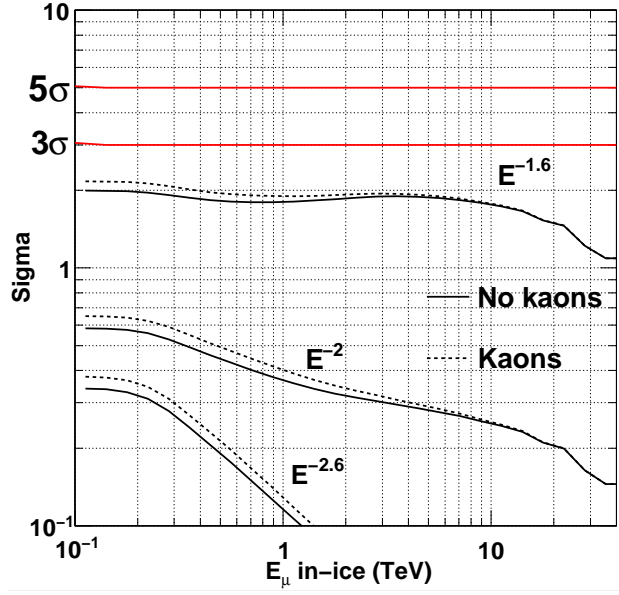, scale=0.45 }
\end{center}
\caption{Statistical significance after 10 years with an energy cut at $E_{\mu}$ of excess muon events due to a point-like PeVatron with flux at 1 TeV of $10^{-11}\,\rm{TeV^{-1}\,cm^{-2}\, s^{-1}}$, for $E^{-1.6}$, $E^{-2}$ and $E^{-2.6}$, with kaons included (dotted) and without kaons (solid). }
\label{fig:pevsig}
\end{figure}

\begin{figure}[hp]
\begin{center}
\epsfig{file=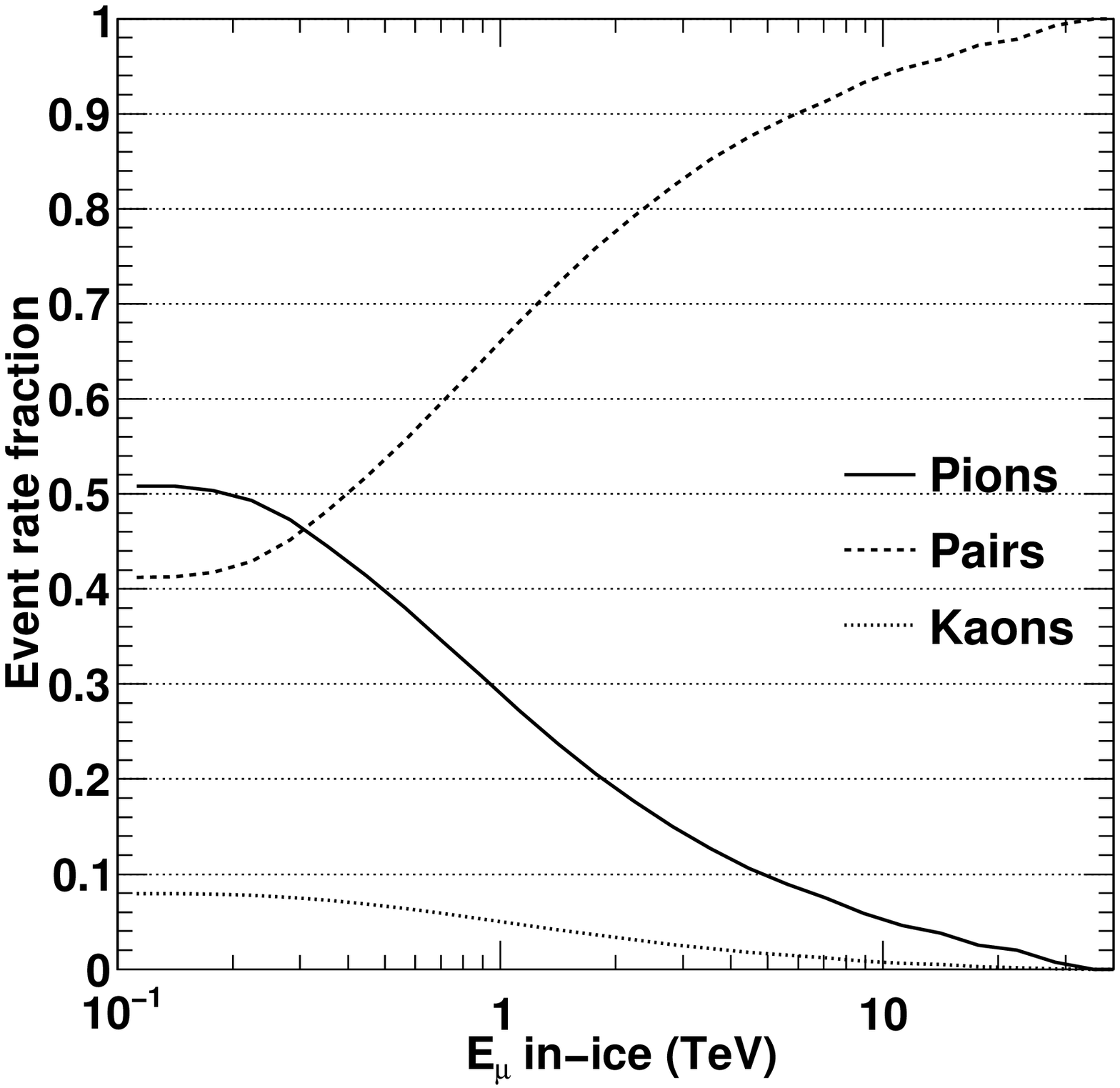, scale=0.45 }
\end{center}
\caption{Fraction of total signal events from a PeVatron due to pionic, kaonic and pair muons for $E^{-1.6}$.  }
\label{fig:pevfrac16}
\end{figure}

\begin{figure}[hp]
\begin{center}
\epsfig{file=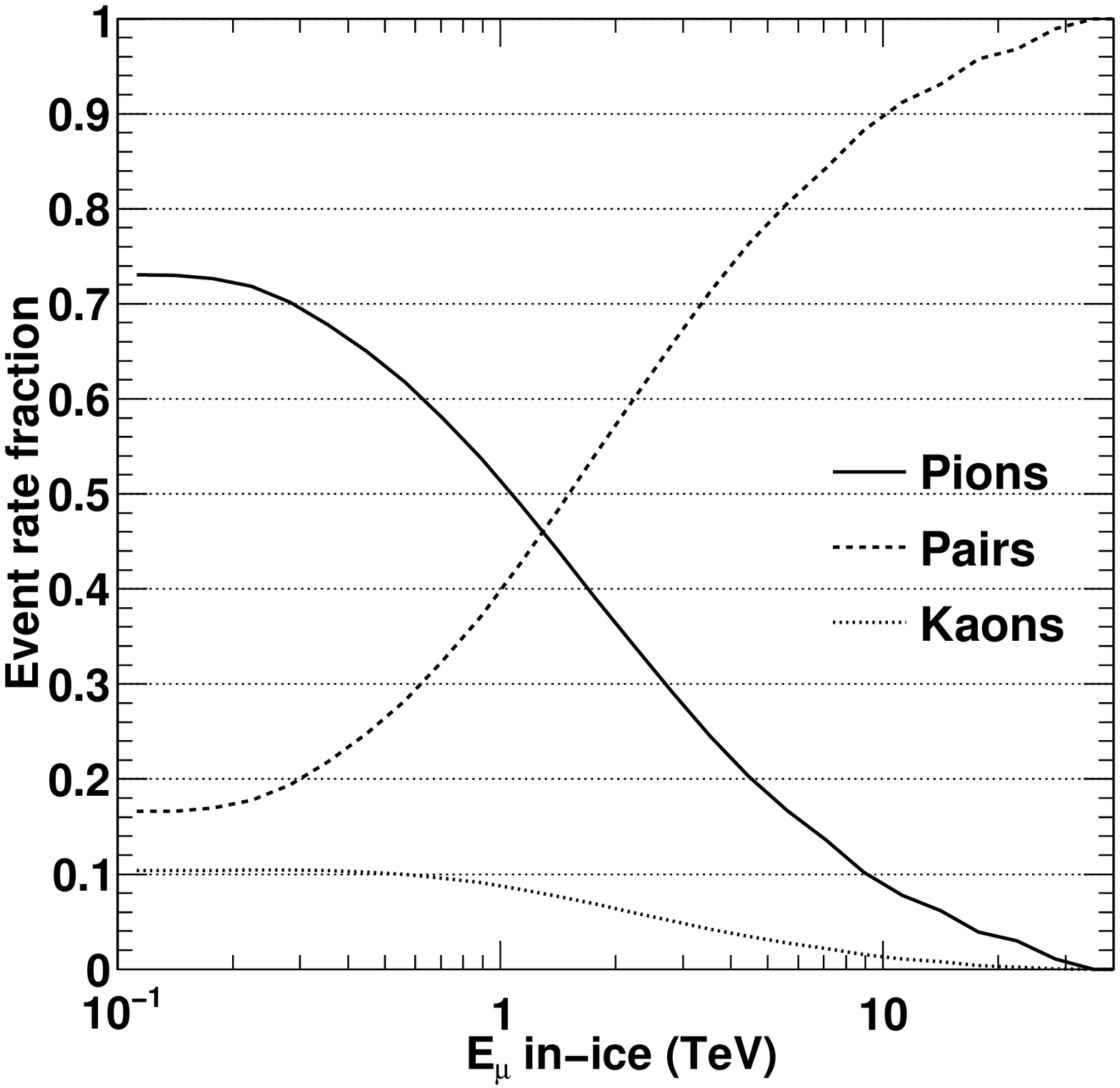, scale=0.45 }
\end{center}
\caption{Fraction of total signal events from a PeVatron due to pionic, kaonic and pair muons for $E^{-2}$.  }
\label{fig:pevfrac2}
\end{figure}

\begin{figure}[hp]
\begin{center}
\epsfig{file=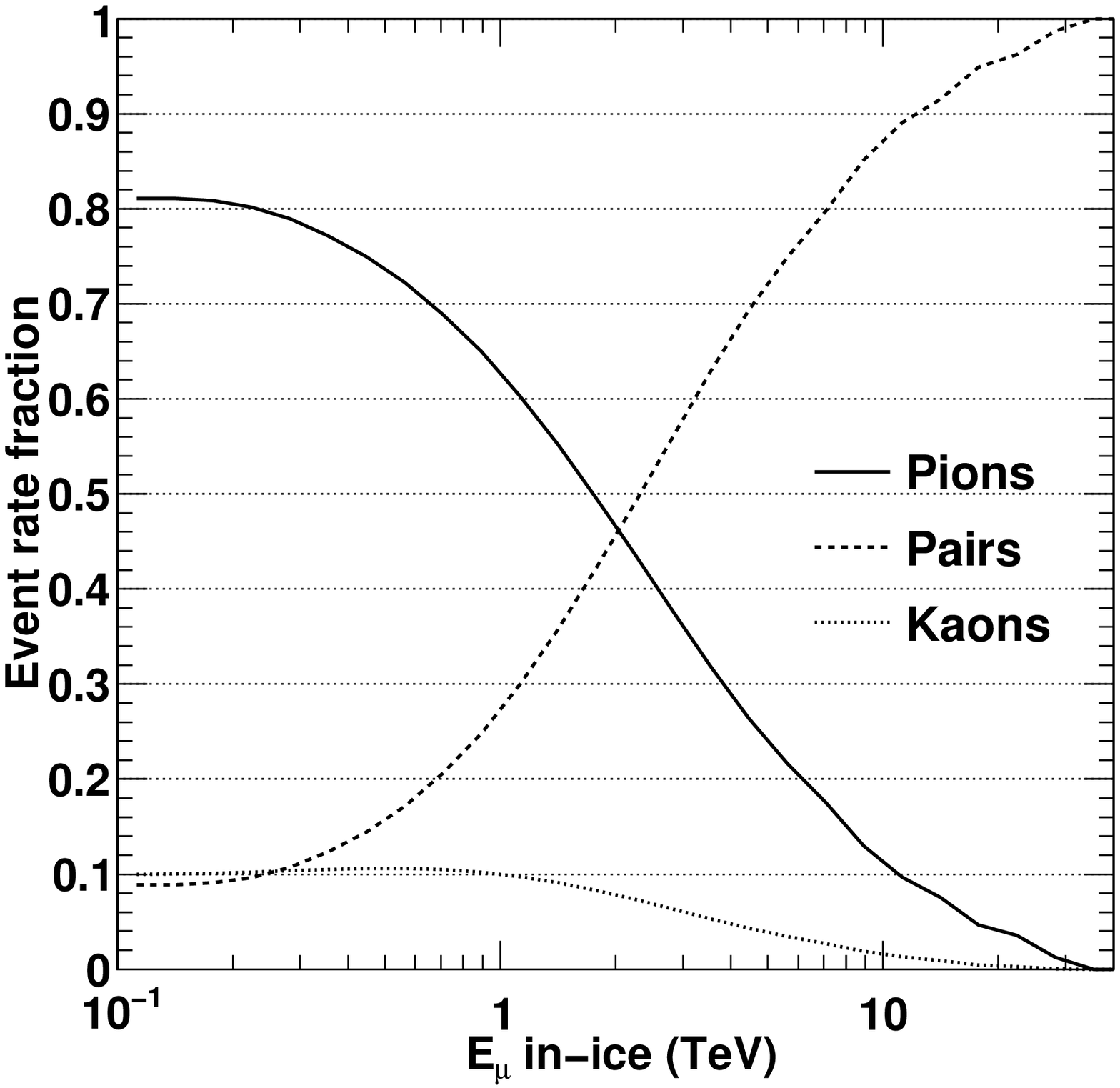, scale=0.45 }
\end{center}
\caption{Fraction of total signal events from a PeVatron due to pionic, kaonic and pair muons for $E^{-2.6}$.  }
\label{fig:pevfrac26}
\end{figure}

{\it Point-like PeVatrons}: A point source of cosmic rays with energies up to the knee at 3\,PeV is expected to be a source of gamma-rays with energies up to $\sim \!E_{\rm{knee}}/10=300\,\rm{TeV}$.  To model a generic PeVatron, we take a source with normalization $K_{\gamma}=1\times10^{-11}\,\rm{TeV^{-1}\,cm^{-2}\, s^{-1}}$ at $1\,\rm{TeV}$ and $E_{\rm{max}}=300\,\rm{TeV}$. We assume an IceCube angular resolution of $0.5^{\circ}$ and a negligible source diameter. The muon flux from such a source, with incident gamma-ray spectral indices of -1.6, -2, and -2.6, is shown in Figs.~\ref{fig:pevflux16},~\ref{fig:pevflux2}, and~\ref{fig:pevflux26}, respectively. The event rates per year are shown in Fig.~\ref{fig:pevev}. The sources are assumed to be at a zenith angle of $30^{\circ}$ but the final statistical significance (Fig.~\ref{fig:pevsig}) varies very little with zenith angles from $0^{\circ}$ to $45^{\circ}$. This is due to the fact that while the mesonic muon fluxes (including the background) at high energy are proportional to $1/\cos\theta$, the distance from the surface to the detector varies by the same factor and so the surface threshold is higher for sources with large zenith angle. Since the maximum significance is always at threshold, Fig.~\ref{fig:pevnorm} shows the minimum normalization as a function of spectral index necessary to observe a point-like PeVatron at $5\,\sigma$ (P$=2.9\times 10^{-7}$) and $3\,\sigma$ (P$=1.3\times 10^{-3}$) after 10 years' time without any cut on muon energy. From these numbers we determine that observing point-like PeVatrons with spectra steeper than $E^{-1.8}$ will require normalizations at 1\,TeV of approximately $3-8\times10^{-11}\,\rm{TeV^{-1}\,cm^{-2}\, s^{-1}}$ (Figure~\ref{fig:pevnorm}). Finally, in Figs.~\ref{fig:pevfrac16},~\ref{fig:pevfrac2}, and~\ref{fig:pevfrac26} we show the fractional contribution to the total signal event rate from each component considered in Section~\ref{sec:fluxes}. 

{\it Extended PeVatrons}: If a very strong extended source exists in the southern sky, such as a large molecular cloud complex powered by several cosmic ray beams, how large can it be to still be detectable by IceCube? We find that a circular source with normalization $10^{-9}\,\rm{TeV^{-1}\,cm^{-2}\, s^{-1}}$ at $1\,\rm{TeV}$ and spectral index $E^{-2}$ gives $5\,\sigma$ after 10 years with a diameter of  $11.5^{\circ}$ (not including kaons) or $13^{\circ}$ (including kaons).  $3\,\sigma$ is reached after 10 years by a source with diameter $19.5^{\circ}$ (not including kaons) or $22^{\circ}$ (including kaons).

\subsection{Vela Jr. }

\begin{figure}[hp]
\begin{center}
\epsfig{file=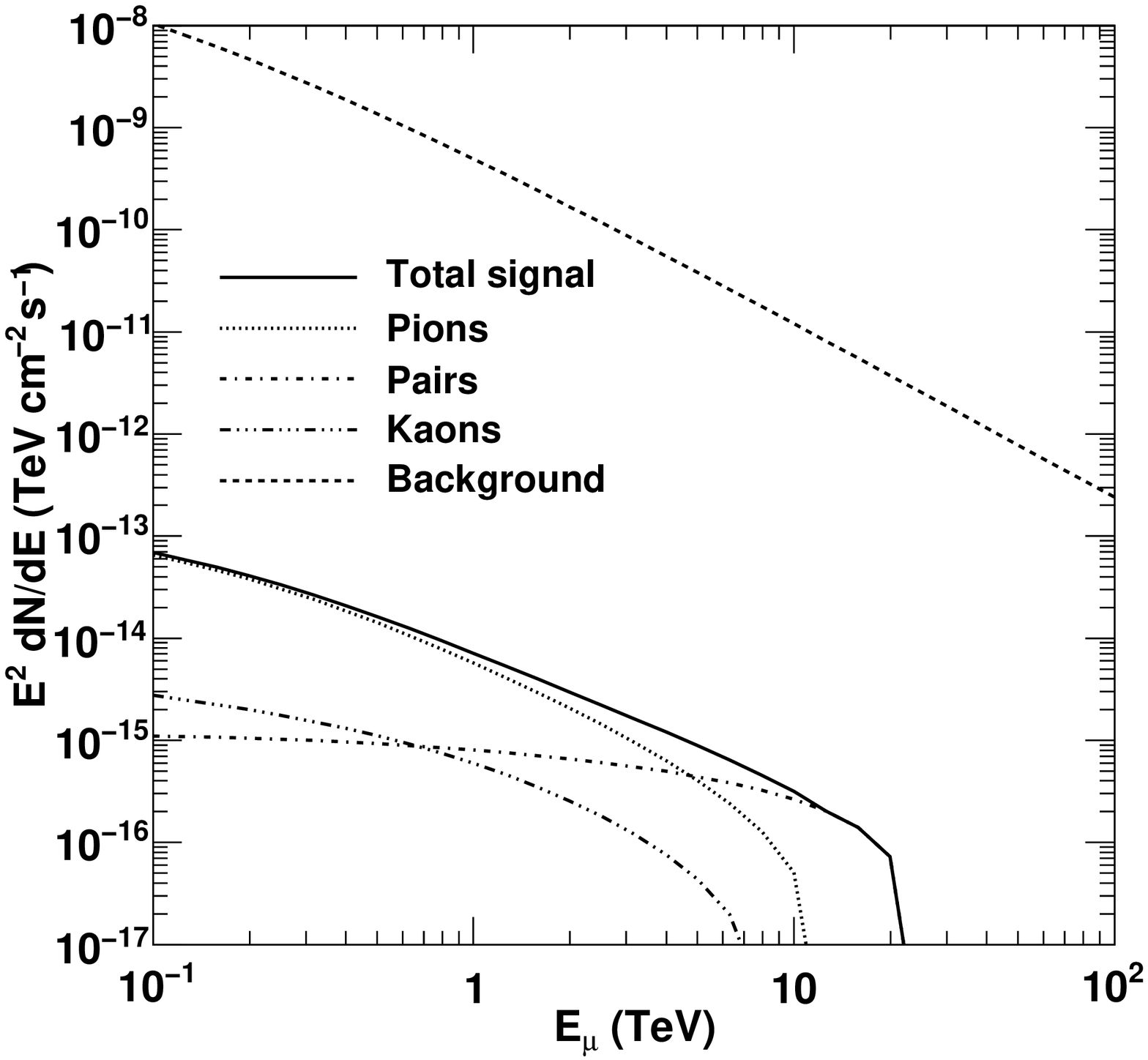, scale=0.45 }
\end{center}
\caption{Differential flux from Vela Jr. with $E_{\rm{max}}=50$ TeV at the source }
\label{fig:velajrflux}
\end{figure}

\begin{figure}[ht]
\begin{center}
\epsfig{file=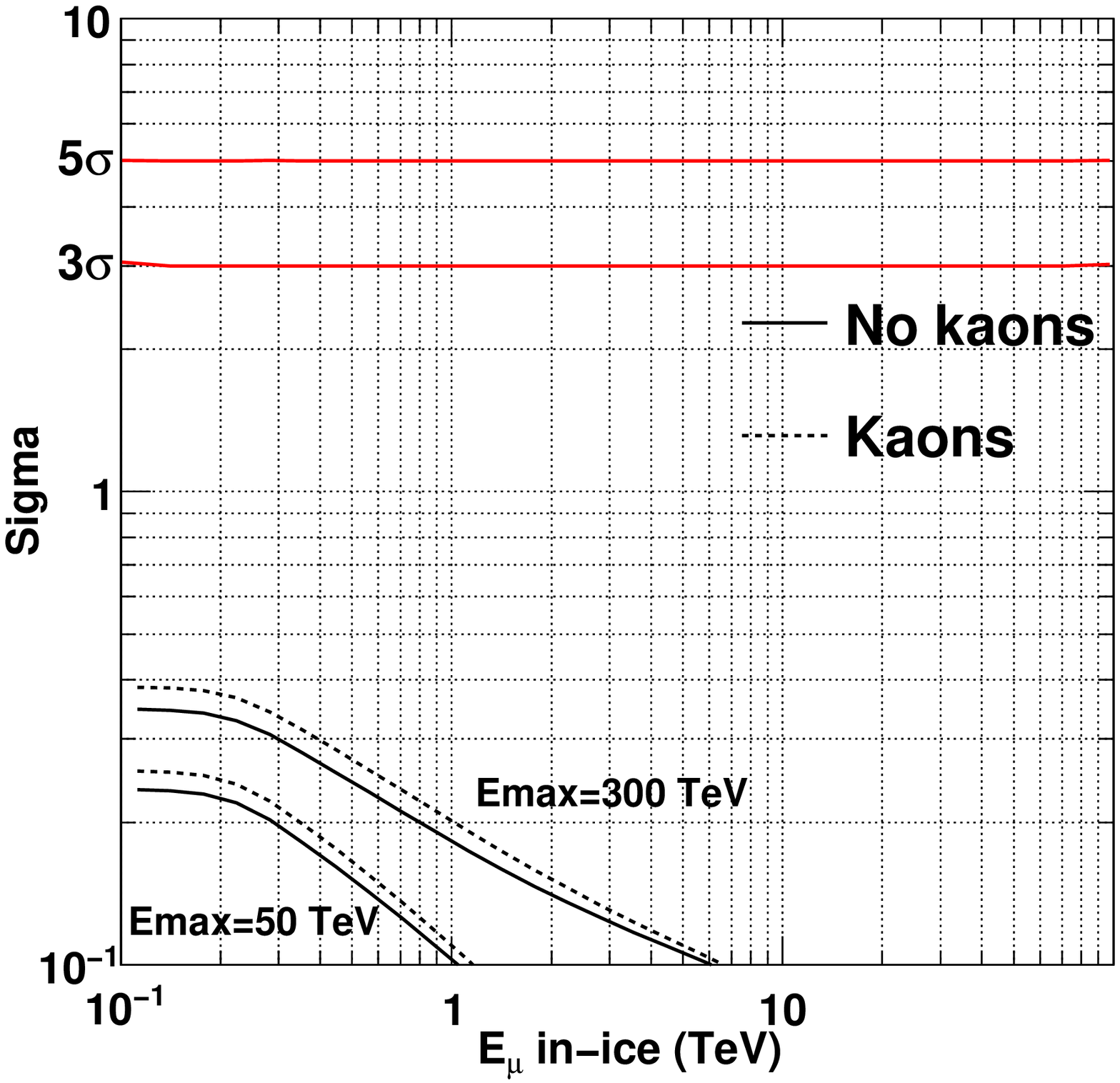, scale=0.5 }
\end{center}
\caption{Statistical significance of Vela Jr. after 10 years for $E_{\rm{max}}=50 \,\rm{TeV},\, 300\,\rm{TeV}$, with kaons included (dotted) and without kaons (solid).  }
\label{fig:velajrsig}
\end{figure}

The brightest TeV source in the Southern Hemisphere, Vela Jr. is a $\sim\!700$-year-old supernova remnant almost directly in front of the Vela Pulsar, at a zenith angle of $\sim\!45^{\circ}$. H.E.S.S. observations of the remnant reveal a $2^{\circ}$ diameter shell emitting gamma-rays with a normalization at 1 TeV of $1.89\times10^{-11} \,\rm{TeV^{-1} \,cm^{-2}\, s^{-1}}$ and an $E^{-2.2}$ spectral index\,\cite{HESS_VelaJr}. H.E.S.S. does not measure above 20 TeV and since it is not known at what energy the spectrum cuts off, we show the differential flux in Fig.~\ref{fig:velajrflux} with $E_{\rm{max}}=50 \,\rm{TeV}$ and the significance after 10 years in Fig.~\ref{fig:velajrsig} with $E_{\rm{max}}=50 \,\rm{TeV}\,\rm{and}~ 300\,\rm{TeV}$. Although Vela Jr. is a strong source, its large diameter means that the background is large and that detection is unlikely.    

\subsection{Other Sources }
We also considered the detection of extragalactic gamma-ray sources such as active galaxies and gamma-ray bursts. Due to the showering, the flux of muons is strongly dependent on the upper energy cutoff of the gamma-ray spectrum. Unfortunately, the existence of extragalactic background photons means that an extragalactic source must be extremely close for photons of energy greater than $\sim\!10\,\rm{TeV}$ to reach the Earth\,\cite{magic_abs}, so even sources that have large fluxes at TeV energies will not be detectable using shower-generated muons. We therefore do not consider extragalactic sources here. 

Finally, the possibility exists of detecting periodic gamma-ray sources such as LS5039, using the periodic modulation of the total event rate to establish the existence of a source. Unfortunately, the known sources are weak and have low enough cutoffs that there will likely be insufficient signal events per period to make a detection. It is also difficult to determine the significance of a periodic source without data. The signal is detected by searching for peaks in the Fourier transform of the total event rate, and the theoretical simulation of real data is done by adding random noise to the mean event rates. However, given the low signal event rates, it is easy for the noise to mask the source frequency, and changing the set of random numbers used greatly affects the height of power spectrum peaks. Hence, we cannot determine an ``average'' sensitivity to a hypothetical periodic source and must wait for actual data whose noise will either allow for the detection of a periodic source or will not. 

\section{Conclusions}

We have calculated the rates and sensitivities of IceCube as a TeV gamma ray observatory.  IceCube, capable of detecting muons of energy $\sim\! 0.1 \, \rm{TeV}$ and above, can observe the presence of muons generated in multi-TeV gamma ray showers, and distinguish these events from background given a sufficiently bright source. While air \c{C}erenkov telescopes are considerably more sensitive, they do not have the ability to monitor large portions of the sky continuously. Even though IceCube has a lower sensitivity to gamma-rays than Milagro, if a TeV-bright transient source occurs in the Southern Hemisphere, IceCube may be the only experiment capable of monitoring it. No additional hardware or software is needed in IceCube beyond its planned design. 

Our results indicate that a standard $E^{-2}$ point source of very-high-energy photons is observable with IceCube provided its flux is greater than several $10^{-11}$ (Figure~\ref{fig:pevnorm}). By contrast, a large emission region composed of many individual PeVatrons is also observable given an extremely strong total flux, approximately $10^{-9} \,\rm{TeV^{-1}\,cm^{-2}\,s^{-1}}$ at 1 TeV for a source of diameter greater than $\sim 13^{\circ}$. 

In all cases, due to the large numbers of muons, the maximum statistical significance is obtained without any cut on muon energy. Since the flux of muons from pair production follows the parent gamma-ray spectrum rather than being a power steeper as in the case of mesonic muons, it might be expected that the greatest sensitivity would be at energies greater than several TeV due to the steepness of the atmospheric background. This does not seem to be the case as the extremely large background of cosmic-ray muons is not sufficiently reduced at higher energies to compensate for the neglecting of many signal muons from pion decay. Therefore, the best search method seems to be to simply maximize the number of signal events by measuring the total number of down-going muons that trigger the detector.

\begin{acknowledgments}
{\footnotesize The authors would like to thank Martin Merck, Teresa Montaruli, Todor Stanev, Juande Zornoza,  and the IceCube Collaboration. This research was supported in part by the National
Science Foundation under Grant No.~OPP-0236449, in part by the
U.S.~Department of Energy under Grant No.~DE-FG02-95ER40896, and in
part by the University of Wisconsin Research Committee with funds
granted by the Wisconsin Alumni Research Foundation. A.K. acknowledges
support by the EU Marie Curie OIF program.}
\end{acknowledgments}

\bibliography{gammashowers}{}

\end{document}